\documentclass[a4paper,11pt]{article}
\usepackage{jcappub} % for details on the use of the package, please
\usepackage{amsmath}
\usepackage{amssymb}
\usepackage{mathrsfs}
\usepackage{bm}
\usepackage{tensor}

\usepackage[T1]{fontenc} % if needed

\usepackage{slashed}
\usepackage{xcolor}

\newcommand{\semibold}[1]{{\fontseries{b}\selectfont{#1}}}

\newcommand{\para}[1]{\par\vspace{2mm}\noindent\semibold{{#1.}---}\ignorespaces} 

\newcommand{\GeV}{\mathrm{GeV}}

\newcommand{\TeV}{\mathrm{TeV}}
\newcommand{\eV}{\mathrm{eV}}
\newcommand{\meV}{\mathrm{meV}}
\newcommand{\second}{\mathrm{s}}
\newcommand{\fb}{\mathrm{fb}}

\newcommand{\Smatter}{S_{\mathrm{m}}}
\newcommand{\LagrangianKernel}{\mathscr{L}}

\newcommand{\LagrangianEM}{\LagrangianKernel_{\text{EM}}}
\newcommand{\LagrangianSM}{\LagrangianKernel_{\text{SM}}}
\newcommand{\LagrangianGal}{\LagrangianKernel_{\text{Gal}}}
\newcommand{\LagrangianDirac}{\LagrangianKernel_{\mathscr{D}}}

\newcommand{\LambdaCC}{\Lambda_{\mathsf{C}}}

\newcommand{\GalileonKernel}{\mathcal{G}}
\newcommand{\GalileonOperator}[1]{\GalileonKernel_{{#1}}}

\newcommand{\DiracOperator}{\mathscr{D}}

\newcommand{\e}[1]{\mathrm{e}^{{#1}}}
\renewcommand{\d}{\mathrm{d}}
\newcommand{\grad}{\nabla}

\newcommand{\vect}[1]{\bm{\mathrm{{#1}}}}

\newcommand{\Mp}{M_{\mathrm{P}}}
\newcommand{\im}{\mathrm{i}}

\newcommand{\ScalarEpsilon}{\epsilon_\phi}
\newcommand{\ScalarEpsilonDot}{\dot{\epsilon}_\phi}
\newcommand{\ScalarEpsilonDDot}{\ddot{\epsilon}_\phi}
\newcommand{\cEM}{c_{\text{EM}}}
\newcommand{\meff}{m_{\text{eff}}}
\newcommand{\meffDot}{\dot{m}_{\text{eff}}}
\newcommand{\omegaeff}{\omega_{\text{eff}}}
\newcommand{\thetafast}{\theta_{\text{fast}}}
\newcommand{\thetafastDot}{\dot{\theta}_{\text{fast}}}
\newcommand{\thetafastDDot}{\ddot{\theta}_{\text{fast}}}
\newcommand{\thetaslow}{\theta_{\text{slow}}}

\newcommand{\TEM}{T_{\text{EM}}}
\newcommand{\Tgrav}{T_{\text{grav}}}
\newcommand{\rhoGal}{\rho_{\text{Gal}}}

\newcommand{\rhoTwo}{\rho_{\GalileonOperator{2}}}
\newcommand{\rhoThree}{\rho_{\GalileonOperator{3}}}

\newcommand{\kphys}{k_{\text{phys}}}

\definecolor{SussexCobaltBlue}{HTML}{1d4289}
\definecolor{SussexDeepAquamarine}{HTML}{007a78}
\definecolor{SussexPowderBlue}{HTML}{7da1c4}
\definecolor{SussexCornYellow}{HTML}{f2c75c}
\definecolor{SussexChinaRose}{HTML}{be84a3}
\definecolor{SussexBurntOrange}{HTML}{dc582a}

\hypersetup{colorlinks=true,citecolor=SussexDeepAquamarine,linkcolor=SussexCobaltBlue,urlcolor=SussexBurntOrange}

\title{\boldmath Constraints on a cubic Galileon disformally
coupled to Standard Model matter}

\author{Michaela G. Lawrence,}
\author{David Seery}
\author{and Christian T.~Byrnes}

\affiliation{Astronomy Centre, University of Sussex, Falmer, Brighton BN1 9QH, UK}

\emailAdd{M.G.Lawrence@sussex.ac.uk}
\emailAdd{D.Seery@sussex.ac.uk}
\emailAdd{C.Byrnes@sussex.ac.uk}

\abstract{We consider a disformal coupling between Standard Model matter
and a cubic Galileon scalar sector,
assumed to be a relict
of some other physics that solves the cosmological constant problem
rather than a solution in its own right.
This allows the energy density carried by the Galileon scalar
to be sufficiently small that it evades stringent constraints from
the integrated Sachs--Wolfe effect, which otherwise rules out
the cubic Galileon theory.
Although the model with disformal coupling does not exhibit
Vainshtein screening,
we show there is a cosmological
`screening-like' phenomenon in which
the energy density carried by the Galileon scalar is
suppressed during matter domination
when the quadratic and cubic Galileon operators are
both relevant.
We obtain the explicit
3+1 form of Maxwell's equations in the presence of the disformal
coupling,
and the wave equations that govern electromagnetic waves.
The disformal coupling is known to generate a small mass
that modifies their velocity of propagation.
We use the WKB approximation to study
electromagnetic waves in this theory and show that,
despite remarkable recent constraints from the LIGO/Virgo
observatories that restrict the difference in propagation velocity
between electromagnetic and gravitational radiation to roughly
1 part in $10^{15}$, the disformal coupling is too weak to be
constrained by events such as GW170817
or by the dispersion of electromagnetic
radiation at different wavelengths.}

\begin{document}
\maketitle
\flushbottom

\section{Introduction}
\label{sec:intro}
For some time,
observational constraints on the Hubble rate (both direct and indirect)
have yielded strong evidence for a ``dark energy'' sector causing the expansion rate
to accelerate since redshift $z \approx 0.5$.
Despite significant effort, the nature of this sector remains largely
unknown.
It may well imply new forces that effectively produce long-range
gravitational repulsion.
These forces would necessarily couple to Standard Model matter and
therefore can
be studied using a combination of
terrestrial, astrophysical and cosmological measurements.
For a recent review, see Ref.~\cite{Brax:2020hkc}.

If one or more new forces \emph{are} present, the minimal possibility
is that they are mediated by a scalar field.
Even if this is not the case and the intermediate field transforms
in a higher-dimensional representation of the Lorentz group,
each physical polarization will act like a scalar field---%
but
perhaps with restricted couplings determined by the representation.
By constraining the different ways that presently-undetected
scalar fields can couple to the Standard Model, we can hope
to place indirect constraints on the unknown dark energy sector.

\para{Universal couplings}
To preserve the weak equivalence principle, any new scalar fields
should couple to all forms of matter in the same way.
What are the possibilities?
At linear level, we can write two universal,
diffeomorphism-invariant couplings
for a scalar field $\phi$:
first, $\phi T / M$
where $T = g^{\mu\nu} T_{\mu\nu}$ is the trace of the
energy--momentum tensor
$T_{\mu\nu} \equiv (-2/\sqrt{-g}) \delta \Smatter / \delta g^{\mu\nu}$,
$\Smatter$ is the matter action,
and $M$ is a mass scale characterizing the strength of the
force;
and second, $(\partial_\mu \phi \partial_\nu \phi) T^{\mu\nu} / M^4$
with the same meanings for $T_{\mu\nu}$ and $M$.

We now wish to promote these interactions into a nonlinear completion.
Note that if the constituent species in $\Smatter$
are to obey the weak equivalence principle
then they should couple to a single metric $\tilde{g}_{\mu\nu}$,
even if this is not the metric
$g_{\mu\nu}$ used to build the gravitational sector.
Then the nonlinear interaction must be
$\Smatter = \Smatter(\tilde{g}^{\mu\nu}, \Psi)$,
where $\Psi$ stands schematically for the different
species of matter fields and
$\tilde{g}_{\mu\nu}$ is related to $g_{\mu\nu}$
by a Bekenstein transformation~\cite{Bekenstein92,Bekenstein:2004ne},
\begin{equation}
	\tilde{g}_{\mu\nu} =
	C(\phi) g_{\mu\nu}
	+ \frac{D(\phi)}{M_D^4} \partial_\mu \phi \partial_\nu \phi .
	\label{eq:bekenstein-xfm}
\end{equation}
We describe $\tilde{g}_{\mu\nu}$ as the ``Jordan frame metric''.
The $C$-term is a conformal transformation of $g_{\mu\nu}$;
in contradistinction, the $D$-term is called \emph{disformal}.
To recover the linear interactions
written above we should expand the $C$- and $D$-functions in
Taylor series,
$C = 1 + \phi/M_C + \cdots$
and
$D = 1 + \cdots$,
and work to lowest order
in $1/M_C$ or $1/M_D$,
as appropriate.
There is no expectation that the mass scales characterizing
the conformal and disformal couplings will coincide.
Bekenstein showed that
this procedure
yields the most general interaction between $\phi$
and matter
that respects causality and the
weak equivalence principle~\cite{Bekenstein92}.

\para{Constraints on $M_D$}
If $M_C$ and $M_D$ are sufficiently large,
so that any new forces are weak, then the linearized interactions
will dominate.
The linear conformal coupling
$\phi T / M_C$
will generate a number of complicated interactions,
depending on the vertices already present in $\Smatter$.
Generically, however, any massive species appearing in $\Smatter$
will become endowed with a Yukawa interaction
whose coupling constant depends on the particle mass $m$.
It follows that
except for very light species, the $\phi$-mediated
force will be dominated at low momentum transfer by
Yukawa exchange,
which yields a $1/r^2$ force law with exponential cutoff
$\e{-mr}$.
Such Yukawa forces are known to be highly
constrained~\cite{Adelberger:2003zx}.
However, this objection is not fatal because
it is now understood that these unwanted
forces can be ``screened'' at the cost
of significant complication in the
self-interactions of $\phi$.
In this paper we do not consider the conformal sector
any further.

The disformal coupling $(\partial_\mu \phi \partial_\nu \phi) T^{\mu\nu} / M_D^4$
is substantially harder to detect.
Because it is derivatively coupled, the amplitude for
$\phi$ exchange with a static, non-relativistic source vanishes.
The classical force generated between such sources must therefore
also vanish: the leading contribution to such fifth-forces
is generated at one-loop level and is highly
suppressed~\cite{Kugo:1999mf,Kaloper:2003yf,Brax:2014vva}.

Nevertheless, attempts have been made to constrain $M_D$.
The strongest of these come from collider phenomenology.
Kaloper gave the approximate lower bound $M_D \gtrsim 200 \, \GeV$
based on unitarity of electron--positron annihilation at LEP~\cite{Kaloper:2003yf}.
It was later shown by Brax \& Burrage that because of cancellations
the cross-section for scalar-mediated
fermion annihilation has an energy dependence
that differs from the estimate used in Ref.~\cite{Kaloper:2003yf}.
This yielded a weaker lower bound $M_D \gtrsim 100 \, \GeV$
from monophoton searches at LHC~\cite{Brax:2014vva}.
Brax, Burrage \& Englert went on to consider
oblique corrections, $Z$ boson phenomenology,
and monophoton, dilepton and monojet events~\cite{Brax:2015hma}.
They
concluded that the strongest constraints came from
monojet searches by the CMS collaboration during LHC Run 1,
which yielded
the refined bound
$M_D \gtrsim 650 \, \GeV$.
Currently, the strongest constraint comes from a dedicated
ATLAS analysis using $37 \, \fb^{-1}$ of LHC data collected
in the period 2015--2016 at centre-of-mass energy
$\sqrt{s} = 13 \, \TeV$.
This yields $M_D \gtrsim 1.2 \, \TeV$~\cite{Aaboud:2019yqu}.

Complementary but weaker
constraints can be obtained from astrophysics and cosmology.
Brax et al.\ studied spectral distortions in
the cosmic microwave background (CMB)
due to variations in the speed of light
induced by the disformal coupling~\cite{Brax:2013nsa}.
Later,
Burrage, Cespedes \& Davis analysed constraints from the
power spectrum of CMB anisotropies in a specific scalar model
with ``quartic Galileon'' self-interactions~\cite{Burrage:2016myt}.
Brax \& Davis and Brax, Davis \& Kuntz derived constraints
from gravitational effects including perhihelion advance,
Shapiro time delay, and inspiral of compact objects~\cite{Brax:2018bow,Brax:2019tcy}.

\para{Gravitational waves}
The advent of multi-messenger astronomy has changed this
picture.
It is now possible to test theories of
modified gravity using observations of gravitational waves,
which have been shown to yield extremely powerful constraints.
In particular, in 2017 the LIGO and Virgo
gravitational wave observatories detected
radiation emitted from the binary neutron star
merger GW170817~\cite{PhysRevLett.119.161101}.
Remarkably, this merger event could be associated with
an electromagnetic counterpart which was interpreted
as a gamma-ray burst.
Assuming the gravitational and electromagnetic radiation
was emitted at nearly the same time,
the observed difference in arrival time $|\Delta t| = (1.74 \pm 0.05) \, \second$
over a path length $\sim 10^{15} \, \second$
implies that the (averaged) propagation speed of gravitational
and electromagnetic waves over their common trajectory
can differ by no more than roughly 1 part in
$10^{15}$.

Many authors have noted that disformal couplings modify the
speed of propagation of electromagnetic waves
relative to gravitational waves.
(We will rederive this important result in~\S\ref{sec:em}.)
Therefore, in principle,
multi-messenger events such as GW170817 offer 
an opportunity to place constraints on such couplings, even from
a single incident.
Indeed,
a large number of proposals for dark energy and modified gravity
have been strongly disfavoured
based on GW170817 alone
because they produce a time difference $|\Delta t|$ that is
unacceptably large~\cite{PhysRevLett.119.251304,PhysRevLett.119.251302,
PhysRevLett.119.251301,PhysRevD.97.061501,PhysRevLett.119.251303}.

\para{Galileon models}
\emph{Galileons} are scalar field models with possibly higher-derivative kinetic terms
that nevertheless yield second-order field equations
due to algebraic cancellations~\cite{Nicolis:2008in}.
Fields of this type are plausible candidates to mediate the new forces
appearing in a dark energy sector.
The model was generalized to curved spacetime (``covariantized'') in Ref.~\cite{Deffayet:2009wt}.
There are five possible Galileon operators
that can appear in the Lagrangian,
of which $\GalileonOperator{1}$ is a linear potential
and $\GalileonOperator{2}$ is the ordinary kinetic term,
\begin{subequations}
\begin{align}
    \label{eq:galileon-one}
    \GalileonOperator{1} & = \phi , \\
    \label{eq:galileon-two}
    \GalileonOperator{2} & = - \frac{1}{2} \grad_\mu \phi \grad^\mu \phi \equiv X , \\
    \label{eq:galileon-three}
    \GalileonOperator{3} & = -2 X \Box \phi , \\
    \label{eq:galileon-four}
    \GalileonOperator{4} & = -2 X \Big[
        2 (\Box \phi)^2 - 2\grad_\mu \grad_\nu \phi \grad^\mu \grad^\nu \phi
        + X R
    \Big] , \\
    \label{eq:galileon-five}
    \GalileonOperator{5} & = -2 X \Big[ (\Box \phi)^3
        - 3 (\grad_\mu \grad_\nu \phi \grad^\mu \grad^\nu \phi) \Box \phi
        + 2 \grad_\mu \grad^\nu \phi \grad_\nu \grad^\rho \phi \grad_\rho \grad^\mu \phi
        - 6 G_{\nu \rho} \grad_\mu \grad^\mu \grad^\nu \phi \grad^\rho \phi
    \Big] .
\end{align}    
\end{subequations}
In these equations, $\grad_\mu$ is the covariant derivative constructed from $g_{\mu\nu}$.

Eqs.~\eqref{eq:galileon-one}--\eqref{eq:galileon-five}
are special cases of the operators studied by Horndeski~\cite{Horndeski:1974wa},
restricted to satisfy a shift symmetry
$\phi \rightarrow \phi + c$ at the level of the action.
(In fact, these operators exhibit a symmetry under the larger \emph{Galilean}
group of transformations $\phi \rightarrow \phi + c + b_\mu x^\mu$.
This symmetry is softly broken in the covariantized model by terms of
order $1/\Mp$
and is therefore restored in the limit $\Mp \rightarrow \infty$
where gravity decouples.)
Note that $\GalileonOperator{1}$ does not spoil the shift symmetry,
at least on a cosmological background,
since it transforms as a
total derivative.
These symmetries restrict the operators that can be generated
by quantum corrections, making the set $\GalileonOperator{1}$ to $\GalileonOperator{5}$
radiatively stable among themselves.
A model including all five operators is said to be \emph{quintic}.
By analogy, if we include all operators except
$\GalileonOperator{5}$ we have a \emph{quartic} model.
If we include all operators except
$\GalileonOperator{4}$ and $\GalileonOperator{5}$ we have a \emph{cubic} model.

A cosmological background spontaneously breaks Lorentz invariance
so that time translations $t \rightarrow t  + c$ are no longer a manifest symmetry.
On these backgrounds,
$\GalileonOperator{4}$ and $\GalileonOperator{5}$ modify the speed of propagation
of gravitational waves.
The conclusion of Refs.~\cite{Ezquiaga:2017ekz,Wang:2017rpx,Sakstein:2015jca}
was that such models are excluded
if
$\phi$ sources late-time acceleration
and is compatible with other cosmological measurements,
because
the time lag $|\Delta t|$ between arrival of gravitational and electromagnetic
radiation from GW170817 is much too large.
Leloup et al.\ extended the same conclusion to a
Galileon model with disformal
couplings~\cite{Leloup:2019fas}.

\para{Outline of this paper}
In this paper we pursue a different but related problem.
The conclusions of Refs.~\cite{Ezquiaga:2017ekz,Wang:2017rpx,
Sakstein:2015jca,Leloup:2019fas}
were driven by the need to
switch on some contribution from $\GalileonOperator{4}$ and $\GalileonOperator{5}$
in order to evade constraints from
the integrated Sachs--Wolfe (``ISW'') effect~\cite{Barreira:2014jha,Brax:2015dma}.
(If CMB--galaxy cross correlations measuring the ISW effect are excluded,
any self-accelerating Galileon model is typically able to satisfy
CMB, BAO and $H_0$ constraints
without tuning the coefficients of the
$\GalileonOperator{i}$~\cite{Renk:2017rzu,Barreira:2014jha}.)
In turn, the large ISW signal arises because
the Galileon field makes a large contribution to the Hubble rate.

This is not the only scenario in which one can imagine a Galileon
scalar sector to arise.
For example, it may not happen that the energy density of the Galileon
field is itself responsible for sourcing late-time acceleration.
Indeed, from one point of view such models are hardly more interesting
as a solution of the cosmological constant problem than simply
taking $\Lambda^4 \approx (10^{-3} \, \eV)^4$
and dispensing with a dynamical component.
This is because a self-accelerated Galileon scenario must usually
take $\Lambda^4 = 0$ at the outset, which is
no more justifiable than
choosing $(10^{-3} \, \eV)^4$
unless we invoke some unknown symmetry that would
make $\Lambda = 0$ a fixed point.
While we do not advocate this position dogmatically, it is worth
bearing in mind.
One might be more willing to tolerate
the choice $\Lambda=0$ required for
a dynamical solution if it could
naturally explain the scale $10^{-3} \, \eV$,
or the redshift associated with the onset of acceleration,
but this does not seem to be the case for Galileon scalars.

In this paper we do not assume that Galileon sector is associated
with a solution to the cosmological constant problem.
It may arise as a vestige of other physics that \emph{is}
associated with the solution, for example as the spin-0
polarization of a massive graviton that somehow degravitates
the vacuum.
Alternatively it may have nothing to do with the cosmological
constant at all.
In either case, our aim is to keep the Galileon 
a subdominant contributor
to the cosmological energy budget.

The question to be resolved is whether a disformal
coupling can be ruled out based on GW170817 alone
(or similar measurements), even without modifications
to the propagation velocity of gravitational waves from
$\GalileonOperator{4}$ and $\GalileonOperator{5}$.
Accordingly we take these operators to be absent.
As explained above,
the resulting cubic model would be ruled out by
measurements of the integrated Sachs--Wolfe effect
if its energy density were significant.
But provided it is subdominant, the model is cosmologically
acceptable.%
    \footnote{The cubic Galileon model has a well-known instability
    causing its energy density to grow at late times~\cite{Chow:2009fm}.
    Ultimately this will set a limit on the period for which our
    model could be a viable effective description.
    We will see in~\S\ref{sec:scalar_dynamics} that there are
    acceptable models for which the onset of the instability
    has not yet occurred.}

We do not study the conformal interaction in this paper
and therefore set the Bekenstein $C(\phi)$-function to unity. 
To go further, note that if the $\phi$ shift symmetry
is unbroken then $D(\phi) = 1$.
Making this choice substantially simplifies the analysis,
but still yields the leading contribution unless
the shift symmetry is strongly broken.
(It is also possible that higher-order interactions involving
$\grad_\mu \phi$ are generated by radiative corrections,
but the usual argument of effective field theory shows
that these will be
subdominant at low energy.)

\para{Summary}
In~\S\ref{sec:em} we derive Maxwell's equations with the inclusion of
a disformal coupling. Parts of this analysis have already been given
in Ref.~\cite{Brax:2013nsa},%
    \footnote{The analysis given in this reference also assumes an
    axion-like coupling to the square of the electromagnetic field
    strength tensor,
    $F_{ab} F^{ab}$, which we do not
    invoke.}
but we repeat them here
in order to fix notation and make our account self-contained.
Because the disformal coupling is wavenumber-dependent,
the resulting electrodynamic equations have similarities
to conventional Maxwell theory in a dispersive medium.
In~\S\ref{sec:scalar_dynamics} we discuss cosmological solutions
of the disformally coupled scalar field.
By solving the Maxwell equations on this background we show
that an evolving bundle of light rays propagating over
cosmological distances is unstable, in principle,
to decay into Galileon particles.
A similar ``tired light'' effect is well-known
in theories of axions and axion-like particles,
including a dark energy ``chameleon''
with appropriate coupling~\cite{Raffelt:1987im,Csaki:2001yk,Burrage:2007ew,Brax:2012ie}.
Where the loss of photons from the bundle is not catastrophic,
there are two key observables.
First, the propagation velocity of electromagnetic waves
differs from those of gravitational waves, as explained above.
Second, because the Maxwell equations
on the scalar field background are dispersive,
such a bundle of light rays would disperse as it
travels over cosmological distances.
We discuss both effects and use them to derive constraints on $M_D$.
Finally, we conclude in~\S\ref{sec:conclusion}.

\para{Notation}
We use metric signature $(-,+,+,+)$ and work in units where
$c = \hbar = 1$.
The reduced Planck mass is defined by
$\Mp^{-2} = 8\pi G$
with numerical value
$\Mp \approx 2.435 \times 10^{18} \, \GeV$.
The scalar field action is
$\LagrangianGal \equiv c_2 \GalileonOperator{2} + c_3 \GalileonOperator{3} / \Lambda^3$,
where
$c_2$ and $c_3$ are Wilson coefficients that
are expected to be of order unity,
and
$\Lambda$ is an energy scale that
determines when the nonlinear operator $\GalileonOperator{3}$
becomes important relative to $\GalileonOperator{2}$.
We describe $\GalileonOperator{2}$ and $\GalileonOperator{3}$
as the linear and nonlinear kinetic terms, respectively.
In the remainder of this paper we absorb
$c_3$ into $\Lambda$ without loss of
generality.
Further,
because there is a scaling symmetry in the
Galileon sector we must fix
$c_2$ in order to break the redundancy~\cite{Barreira:2014jha}.
The role of $c_2$ is therefore only to
select whether the quadratic kinetic
term is \emph{individually} stable or ghostlike,
and
by making a further scaling transformation
we can always arrange that $c_2 = \pm 1$.

With these choices,
on a cosmological background, the Lagrangian
density specializes to
\begin{equation} \label{gal_lagrangian_eq}
    \LagrangianGal(\phi) =
    \frac{c_{2}}{2} \dot{\phi}^{2}
    + \frac{1}{\Lambda^3} \ddot{\phi} \dot{\phi}^{2} .
\end{equation}

\section{Maxwell theory}
\label{sec:em}

In this section we derive Maxwell's equations in the presence of
a disformal coupling.
A version of this analysis was previously given
using manifestly Lorentz-covariant
methods
by Brax et al.~\cite{Brax:2013nsa},
who considered electrodynamics in the presence of a disformal coupling
together with an additional axion-like interaction with the Maxwell term.
(For example, an interaction of this form is known to arise
under change of frame; see Ref.~\cite{Brax:2010uq}.)
In comparison with the discussion there, our analysis does not include
the axion-like coupling.
To make the physical content of the model as transparent as
possible we frame our discussion in terms of the
Maxwell equations and the physical $\vect{E}$ and $\vect{B}$
fields.
We comment further on the relation between our calculations
below.

\subsection{Disformally-coupled electromagnetism}
The action is
\begin{equation}
\label{equation1}
    \mathcal{S}=\int \d^{4} x \; \sqrt{-g} \,
    \bigg[
        \frac{\Mp^{2}}{2} R-\LambdaCC
        + \LagrangianGal(\phi)
    \bigg]
    +
    \int \d^{4} x \; \sqrt{-\tilde{g}}
    \LagrangianSM
    (
        \tilde{g}_{\mu \nu}, \Psi
    ),
\end{equation}
where $\LagrangianSM$ is the Standard Model
Lagrangian density, and the remainder of the notation matches
that used in~\S\ref{sec:intro}.
In particular, $\tilde{g}_{\mu\nu}$ is the Jordan frame metric
and $\Psi$ continues to stand schematically for the different
species of Standard Model particles.
Also, $R = R(g)$ is the Ricci scalar constructed from the
`vanilla' metric $g_{\mu\nu}$
and $\LambdaCC \approx (10^{-3} \, \eV)^4$ is a cosmological constant
that is assumed to drive the observed late-time acceleration
of the expansion $a(t)$.

Eq.~\eqref{equation1}
could equivalently be written in the Jordan frame
by exchanging $R = R(g)$ for $\tilde{R} = \tilde{R}(\tilde{g})$,
the Ricci scalar constructed from the Jordan-frame metric
$\tilde{g}$.
The disformal coupling between $\phi$ and matter would then
become manifest as a derivative coupling between $\phi$
and $\tilde{G}_{\mu\nu}$,
where $\tilde{G}_{\mu\nu} = \tilde{R}_{\mu\nu} - \tilde{R} \tilde{g}_{\mu\nu}/2$ is the ordinary
Einstein tensor.
This approach was adopted in Ref.~\cite{Leloup:2019fas}.

In this paper we focus on the Maxwell term in the Standard Model
Lagrangian.
In the presence of a source 4-current $J^\mu = (\rho, \vect{J})$
this can be written
\begin{equation}
    \label{eq:Maxwell-Lagrangian}
    \frac{\LagrangianEM}{\sqrt{-g}} =-\frac{1}{4} F_{\mu \nu} F^{\mu \nu}+J^{\mu} A_{\mu},
\end{equation}
where, as usual,
\begin{equation}
    \label{eq:Maxwell-tensor}
    F_{\mu \nu}=\grad_{\mu} A_{\nu}-\grad_{\nu} A_{\mu} .
\end{equation}
The contravariant metric corresponding to $\tilde{g}_{\mu\nu}$ is
[c.f. Eq.~\eqref{eq:bekenstein-xfm} with $C=1$]
\begin{equation} \label{eq:inverse_metric}
    \tilde{g}^{\mu \nu} = g^{\mu \nu} - \Lambda_D \partial^{\mu} \phi \partial^{\nu} \phi .
\end{equation}
Like $D$,
the scale $\Lambda_D$ has dimension $[\mathrm{M}^{-4}]$.
It is conventionally written in terms of the Einstein-frame
kinetic energy for the scalar field
$X \equiv - g^{\mu\nu} \partial_\mu \phi \partial_\nu \phi / 2$.
Then it follows that
\begin{equation}
    \label{eq:inv_metric_const}
    \Lambda_D = \frac{M_D^{-4}}{1-2X M_D^{-4}} = \frac{M_D^{-4}}{1-\dot{\phi}^2/M_D^4}.
\end{equation}
The last equality applies only
in the special case of a cosmological background, where $\phi$
depends only on coordinate time $t$. In this equation and subsequently,
an overdot denotes a derivative with respect to $t$.

The integration measures $\sqrt{-g}$ and $\sqrt{-\tilde{g}}$
have a well-known relation~\cite{Bekenstein:2004ne,Bettoni:2013diz},
\begin{equation}
\label{eq:measure}
    \sqrt{-\tilde{g}} = (1 - 2X M_D^{-4})^{1/2}\sqrt{-g} = (1- \dot{\phi}^2/M_D^4)^{1/2}\sqrt{-g}.
\end{equation}
Moreover the Christoffel symbols are related by~\cite{Bettoni:2013diz}
\begin{equation}
    \label{eq:tilde-connection}
    \tilde{\Gamma}^\rho_{\mu \nu} = 
    \Gamma^\rho_{\mu \nu} 
    + \Lambda_D \grad^\rho \phi \grad_\mu \grad_\nu \phi.
\end{equation}
These formulae allow us to exchange covariant derivatives $\grad_\mu$
compatible with the metric $g_{\mu\nu}$
for derivatives $\tilde{\grad}_\mu$ compatible with $\tilde{g}_{\mu\nu}$.

\para{Variational principle}
To derive the Maxwell equations it is most convenient to write the action in
`Schwinger' form.
This is analogous to the Palatini formulation of Einstein gravity,
in which one takes the Riemann tensor
$\tensor{R}{^\lambda_\mu_\nu_\rho} = \tensor{R}{^\lambda_\mu_\nu_\rho}(\Gamma)$
to be constructed from
the connection $\tensor*{\Gamma}{^\lambda_\mu_\nu}$, but without any
assumption regarding the relation between
$\tensor*{\Gamma}{^\lambda_\mu_\nu}$ and $g_{\mu\nu}$.
One then treats $\tensor*{\Gamma}{^\lambda_\mu_\nu}$
and $g_{\mu\nu}$ as independent fields.
After variation with respect to
$\tensor*{\Gamma}{^\lambda_\mu_\nu}$ and $g_{\mu\nu}$,
the Einstein equations for $\tensor{R}{^\lambda_\mu_\nu_\rho}(\Gamma)$
follow from demanding that bulk contributions vanish.
Meanwhile,
demanding that boundary terms vanish
requires $\grad_\mu$ to be compatible with $g_{\mu\nu}$
and therefore determines 
$\tensor*{\Gamma}{^\lambda_\mu_\nu} = \tensor*{\Gamma}{^\lambda_\mu_\nu}(g)$
to be the Levi--Civita connection
via the fundamental theory of Riemannian geometry.

A similar approach due to Schwinger can be applied to the Maxwell Lagrangian.%
    \footnote{See lecture 4 in the notes on quantum field theory by Ludwig Fadeev
    published in Ref.~\cite{Deligne:1999qp}.}
To proceed, replace the Maxwell action~\eqref{eq:Maxwell-Lagrangian} by
\begin{equation}
    \label{eq:Schwinger-action}
    \frac{\LagrangianEM}{\sqrt{-g}} = - \frac{1}{2} (\grad_\mu A_\nu - \grad_\nu A_\mu) F^{\mu\nu}
    + \frac{1}{4} F_{\mu\nu} F^{\mu\nu} + J^\mu A_\mu .
\end{equation}
The field-strength tensor $F_{\mu\nu}$ and the connection $A_\mu$ are to be regarded
as independent,
making the Lagrangian density linear in derivatives as for the Palatini procedure.
Variation with respect to $F^{\mu\nu}$ clearly
reproduces the expected definition of the Maxwell tensor, Eq.~\eqref{eq:Maxwell-tensor}.
Substitution of this result in~\eqref{eq:Schwinger-action}
yields~\eqref{eq:Maxwell-Lagrangian},
and therefore
we conclude that both variational principles are equivalent.

We now introduce the distinction between
$g_{\mu\nu}$ and $\tilde{g}_{\mu\nu}$.
Following the same procedure that
led to~\eqref{eq:Schwinger-action},
we find
\begin{equation}
\label{em_eq_bars}
    \LagrangianEM[\tilde{g}, A, J]
    =
    \sqrt{-\tilde{g}}
    \bigg(
        {-\frac{1}{2}} (\tilde{\grad}_{\mu} A_{\nu}  -\tilde{\grad}_{\nu} A_{\mu}) \tilde{F}^{\mu \nu} 
        + \frac{1}{4} \tilde{F}_{\mu \nu} \tilde{F}^{\mu \nu}
    \bigg)
    +
    \sqrt{-g} \, J^\mu A_\mu
    ,
\end{equation}
where $\tilde{\grad}$ is the covariant derivative constructed from the
connection $\tilde{\Gamma}$ given in Eq.~\eqref{eq:tilde-connection},
and $\tilde{F}_{\mu\nu}$ is the Maxwell tensor built from
$\tilde{\grad}$
and the electromagnetic 4-potential $A_\mu$.

\para{Transformation of the source current}
We are free to express the theory in terms of
whatever frame is most convenient.
In this section our aim is to obtain the Maxwell equations
for the Einstein-frame $\vect{E}$
and $\vect{B}$
fields.
This is useful because eventually it is the
Einstein-frame metric $g_{\mu\nu}$ that will carry
an FRW cosmology.
The Hubble rate for this metric will receive
contributions from the Einstein-frame
$\vect{E}$ and $\vect{B}$ fields, modified by their interactions
with the $\phi$ field.

For this purpose we require the Einstein frame
4-current $J^\mu = (\rho, \vect{J})$
appearing in Eq.~\eqref{em_eq_bars}, whose time
and space components are the
charge density $\rho$ and 3-current $\vect{J}$ respectively.
Although we are working in the Einstein frame,
for practical purposes it is convenient
to express these in
terms of the equivalent \emph{Jordan-frame} quantities,
because it is these that would be measured by an experimentalist
working in a small freely-falling laboratory
in which the influence of gravity and fifth-forces
can be neglected.
To determine exactly how $A_\mu$ couples
to these quantities
we begin with the action for a Dirac spinor $\psi$
coupled to the Jordan-frame metric $\tilde{g}_{\mu\nu}$,
 \begin{equation}
 	\label{eq:DiracAction-JordanFrame}
    \int \d^4 x \; \LagrangianDirac
    =
    \int
    \mathrm{d}^4x \; \sqrt{-\tilde{g}} \,
    \Big[
        {- \bar{\psi}} \slashed{\DiracOperator} \psi + \mathrm{h.c.}
    \Big],
\end{equation}
where
$\slashed{\DiracOperator} \equiv \gamma^a
\tensor{\tilde{e}}{_a^\mu}(\tilde{\grad}_\mu - \im q A_\mu)$
is the gauge- and
diffeomorphism-covariant Dirac operator for a spin-$1/2$
fermion of charge $q$,
$\tensor{\tilde{e}}{_a^\mu}$ is a vierbein for $\tilde{g}_{\mu\nu}$,
and $\bar{\psi} = \psi^\dag \gamma^0$ is the adjoint spinor to $\psi$.
Greek indices $\mu$, $\nu$, {\ldots}
label tensor indices transforming under spacetime coordinate
diffeomorphisms,
whereas Latin indices $a$, $b$, {\ldots}
label indices transforming under the tangent space Lorentz group
$SO(1,3)$.
In particular, the vierbein satisfies
\begin{equation}
    \tensor{\tilde{e}}{_a^\mu} \tensor{\tilde{e}}{^a^\nu} = \tilde{g}^{\mu\nu}
    \qquad\text{and}\qquad
    \tensor{\tilde{e}}{_a^\mu} \tensor{\tilde{e}}{_b_\mu} = \eta_{ab} .
\end{equation}
Finally, the $\gamma^a$ are Dirac matrices transforming
under the tangent space Lorentz group. They satisfy the usual Dirac algebra
$\{ \gamma^a, \gamma^b \} = 2 \eta^{ab} \vect{1}$,
where $\vect{1}$ denotes the identity matrix in the Dirac spinor representation.
The action of the covariant derivative on a spinor can be written
\begin{equation}
    \grad_\mu \psi = \partial_\mu \psi + \frac{1}{8} \omega_\mu^{ab} \gamma_{ab} \psi ,    
\end{equation}
where $\gamma_{ab} \equiv [ \gamma_a, \gamma_b ]$
and $\omega_\mu^{ab}$ is the spin connection.

To identify the charge density $\rho$
and current $\vect{J}$
we break Eq.~\eqref{eq:DiracAction-JordanFrame}
into space and time components.
Notice that after doing so our expressions appear to mix
indices transforming under the $SO(1,3)$ and diffeomorphism
groups, although this appearance is fictitious.
We find
\begin{equation}
\label{eq:DiracLagrangian-Split}
    \frac{\LagrangianDirac}{\sqrt{-g}}
    =
    {-\bar{\psi}} \gamma^0\tilde{\grad}_0 \psi
    - \frac{(1-\dot{\phi}^2/M_D^4)^{1/2}}{a} \bar{\psi} \gamma^i \tilde{\grad}_i \psi
    +\im q
    \Big(
        \bar{\psi} \gamma^0 A_0 \psi
        + \frac{(1-\dot{\phi}^2/M_D^4)^{1/2}}{a} \bar{\psi} \gamma^i A_i \psi
    \Big)
    + \mathrm{h.c.}
    ,
\end{equation}
where $a = a(t)$ is the scale factor
and `h.c.' denotes the Hermitian conjugate of the entire preceding
expression.
Spatial indices $i$, $j$, {\ldots}
should be summed using the spatial part of the
Einstein-frame FRW metric $g_{ij} = a^2 \delta_{ij}$.
Identifying $\tilde{\rho} = \im q \bar{\psi} \gamma^0 \psi + \text{h.c.}$
as the Jordan-frame charge density
and $\tilde{\vect{J}} = \im q \bar{\psi} \gamma^i \psi / a + \text{h.c.}$
as the corresponding 3-current, we conclude
\begin{equation}
    \label{eq:source-disformal-xfm}
	\rho = \tilde{\rho}
	\qquad\text{and}\qquad
	\vect{J} = (1 - \dot{\phi}^2/M_D^4)^{1/2} \tilde{\vect{J}} .
\end{equation}

Returning to Eq.~\eqref{em_eq_bars},
expressing all quantities in terms of the Einstein frame,
and using~\eqref{eq:source-disformal-xfm} for $J^\mu$, we obtain
\begin{equation}
\begin{split}
    \frac{\LagrangianEM}{\sqrt{-g}}
    =
    \mbox{}
    &
    {-\ScalarEpsilon}
    (\partial_{0} A_{i}-\partial_{i}A_0) F^{0i}
    - \frac{1}{2\ScalarEpsilon} (\partial_{i} A_{j}-\partial_{j} A_{i}) F^{ij}
    + \frac{\ScalarEpsilon}{2} F_{0i} F^{0i}
    + \frac{1}{4\ScalarEpsilon} F_{ij} F^{ij}
    \\
    &
    + \tilde{\rho} A_0 + \frac{1}{\ScalarEpsilon} \tilde{J}^i A_i .
\end{split}
\end{equation}
We have used~\eqref{eq:inv_metric_const}
and defined the quantity $\ScalarEpsilon$ to satisfy
\begin{equation}
    \ScalarEpsilon
    \equiv
    \bigg( 1 - \frac{\dot{\phi}^2}{M_D^4} \bigg)^{-1/2} .
\end{equation}
To proceed
we introduce the Einstein-frame $\vect{E}$ and $\vect{B}$ fields using the conventional
definitions%
    \footnote{An observer moving in spacetime with 4-velocity $u_\mu$, normalized
    (with our metric convention) so that $u_\mu u^\mu = -1$,
    would observe electric and magnetic fields defined by
    \begin{subequations}
    \begin{equation}
        E_\mu = u^\nu F_{\mu\nu}
    \end{equation}
    and
    \begin{equation}
        B_\mu = - \frac{1}{2} \epsilon_{\mu\nu\rho\sigma} u^\sigma F^{\nu\rho} ,
    \end{equation}
    \end{subequations}
    where $\epsilon_{\mu\nu\rho\sigma}$
    is the four-dimensional Levi--Civita tensor
    normalized so that $\epsilon_{0123}= \sqrt{-g}$.
    An observer comoving with the cosmological expansion has
    $u^\mu = (1, \vect{0})$,
    from which Eqs.~\eqref{eq:ElectricFieldDef} and~\eqref{eq:MagneticFieldDef}
    follow. See, e.g., Ref.~\cite{Tsagas:2004kv}.
    }
\begin{subequations}
\begin{align}
    \label{eq:ElectricFieldDef}
    a E_{i}&=-F_{0 i}, \\
    \label{eq:MagneticFieldDef}
    a B_{i}&=\frac{1}{2} \epsilon_{i j k} F^{j k} ,
\end{align}
\end{subequations}
where $\epsilon_{ijk}$ is the covariant Levi-Civita tensor;
its components take the values $\pm \sqrt{-g}$.
In terms of these fields the action can be rewritten
\begin{equation}
\label{eq:final-maxwell-action}
\begin{split}
    \frac{\LagrangianEM}{\sqrt{-g}}
    =
    \mbox{}
    &
    {-\frac{\ScalarEpsilon}{a}} \vect{E} \cdot \frac{\partial}{\partial t} (a^2 \vect{A})
    + \frac{\ScalarEpsilon}{a} \phi \grad\cdot\vect{E}
    - \frac{1}{\ScalarEpsilon} \vect{A} \cdot \grad \times \vect{B}
    - \frac{\ScalarEpsilon}{2} \vect{E}^2
    + \frac{1}{2 \ScalarEpsilon} \vect{B}^2
    \\
    &
    -
    \tilde{\rho} \phi
    +
    \frac{a^2}{\ScalarEpsilon} \tilde{\vect{J}} \cdot \vect{A}
    ,
\end{split}
\end{equation}
where we have integrated by parts,
dropped boundary terms at spatial infinity,
and used that $D = 1/M_D^4$ is spatially independent
in our model.
(If $D$ has spatial dependence then the action has a more complicated formulation.)
We have also dropped explicit summation over spatial indices
in favour of ordinary dot and cross products in a three-dimensional Euclidean metric.
In this 3+1 split the electromagnetic 4-potential
satisfies $A^\mu = (\phi, \vect{A})$, where $\phi$ and $\vect{A}$
are the three 3-dimensional scalar and vector potential respectively.

\subsection{Maxwell's equations}

Maxwell's equations can be obtained from Eq.~\eqref{eq:final-maxwell-action}
by variation with respect to $\phi$, $\vect{A}$, $\vect{E}$ and $\vect{B}$.
Of these, $\phi$ and $\vect{B}$ enter the action in terms that do not involve
time derivatives, and therefore produce constraints rather than dynamical equations.
The variations of $\vect{A}$ and $\vect{E}$ produce the evolution equations
of the theory.

\para{Constraints}
To see this in detail, first perform the variation with respect to $\phi$.
This yields Gauss' law,
\begin{subequations}
\begin{equation}
    \label{eq:GaussConstraint}
    \frac{1}{a} \grad \cdot \vect{E}
    = \frac{\tilde{\rho}}{\ScalarEpsilon} .
\end{equation}
By analogy with the usual Maxwell equation
for a medium with permittivity $\epsilon$,
$\grad \cdot \vect{E} = \rho / \epsilon$,
we see that the electric field
responds to charge density
as if it were immersed within a medium
with electric constant $\ScalarEpsilon = (1 - D \dot{\phi}^2)^{-1/2}$.
However, we will see that this analogy cannot be extended
to all the Maxwell equations.
Meanwhile, variation with respect to $\vect{B}$
enforces conservation of magnetic flux,
\begin{equation}
    \label{eq:FluxConstraint}
    \vect{B} = \grad \times \vect{A}
    \quad \Rightarrow \quad
    \frac{1}{a} \grad \cdot \vect{B} = 0 .    
\end{equation}
As expected, both Eqs.~\eqref{eq:GaussConstraint} and~\eqref{eq:FluxConstraint}
are constraints.

\para{Dynamical equations}
The remaining Maxwell equations follow from variation with respect to
$\vect{A}$ and $\vect{E}$.
The $\vect{A}$ variation yields Amp\`{e}re's circuital law,
\begin{equation}
    \label{eq:AmpereLaw}
    \frac{1}{a}\grad \times \vect{B} =
    a \tilde{\vect{J}}
    +
    \frac{\ScalarEpsilon}{a^2} \frac{\partial}{\partial t}
    \Big( a^2 \ScalarEpsilon \vect{E}
    \Big) .
\end{equation}
By comparison, the form of this law in a medium with fixed
electric and magnetic constants $\epsilon$ and $\mu$
would be
$\grad \times \vect{B}
= \mu \vect{J} + \mu \epsilon \dot{\vect{E}}$.
Therefore, apparently, there is no assignment of
electric and magnetic constants that would maintain
the analogy of the scalar condensate as a
dielectric medium,
as can be done for the gravitational coupling~\cite{Plebanski:1959ff}.

Finally, variation with respect to $\vect{E}$
yields Faraday's law of induction
\begin{equation}
    \label{eq:FaradayLaw}
    \vect{E} = - \frac{1}{a} \grad \phi - \frac{1}{a} \frac{\partial}{\partial t} (a^2 \vect{A})
    \quad \Rightarrow \quad
    \frac{1}{a} \grad \times \vect{E} = - \frac{1}{a^2} \frac{\partial}{\partial t} (a^2\vect{B}) .
\end{equation}
\end{subequations}
This is not modified by the disformal coupling.

Relations equivalent to~\eqref{eq:GaussConstraint}--\eqref{eq:FaradayLaw}
were given in Ref.~\cite{Brax:2013nsa},
although because this reference worked in terms of a covariant formalism
they were not broken out into separate Maxwell equations.
    
\subsection{Electromagnetic waves}
\label{sec:EMWaves}

\para{The wave equation}
Our interest lies in the propagation of electromagnetic
radiation from cosmological distances, and for this
purpose we require an equation governing
electromagnetic waves.
After specializing to the vacuum case,
a suitable equation for the magnetic field $\vect{B}$
can be obtained by taking the curl of Amp\`{e}re's
law, Eq.~\eqref{eq:AmpereLaw},
and substituting the time derivative of Faraday's
law~\eqref{eq:FaradayLaw}
to eliminate $\grad \times \dot{\vect{E}}$.
The wave equation resulting from this procedure is
\begin{subequations}
\begin{equation}
    \label{eq:BWaveEquation}
    \ddot{\vect{B}}
    + \bigg( 5 H + \frac{\ScalarEpsilonDot}{\ScalarEpsilon} \bigg) \dot{\vect{B}}
    + \bigg(
        2 (3 H^2 + \dot{H})
        +
        2 H \frac{\ScalarEpsilonDot}{\ScalarEpsilon}
    \bigg) \vect{B}
    -
    \frac{1}{a^2 \ScalarEpsilon^2} \grad^2 \vect{B}
    = 0 .
\end{equation}
The coupling to gravity has been well-studied~\cite{Turner:1987bw,Tsagas:2004kv}.
Both gravitational effects and the disformal coupling generate
a soft mass term that does not spoil gauge invariance.
In particular note that this wave equation describes a coupled
electric and magnetic oscillation of fixed frequency, not
a free magnetic field.
Therefore, despite the large friction term $5H$ appearing in~\eqref{eq:BWaveEquation},
the energy density $\sim \vect{B}^2$ of
a free magnetic field still redshifts at the rate
$1/a^4$ expected for radiation.
The same applies for a free electric field.
See, e.g., Ref.~\cite{Barrow:2011ic}.

Once a solution for $\vect{B}$ is known, it can be used to generate
a solution for $\vect{E}$
via Eq.~\eqref{eq:FaradayLaw}.
Alternatively, $\vect{E}$ can be solved directly using
\begin{equation}
    \label{eq:EWaveEquation}
    \ddot{\vect{E}}
    + \bigg( 5 H + 3 \frac{\ScalarEpsilonDot}{\ScalarEpsilon} \bigg) \dot{\vect{E}}
    + \bigg(
        2(3H^2 + \dot{H})
        +
        7 H \frac{\ScalarEpsilonDot}{\ScalarEpsilon}
        +
        \frac{\ScalarEpsilonDot^2}{\ScalarEpsilon^2}
        +
        \frac{\ScalarEpsilonDDot}{\ScalarEpsilon}
    \bigg) \vect{E}
    -
    \frac{1}{a^2 \ScalarEpsilon^2} \grad^2 \vect{E}
    = 0 .
\end{equation}
\end{subequations}
Electromagnetic waves governed by Eqs.~\eqref{eq:BWaveEquation}
or~\eqref{eq:EWaveEquation}
do not propagate with velocity $c=1$.
Neglecting the effect of the 
mass term, the
`sound speed' determined by the
ratio of spatial to temporal kinetic
terms is
\begin{equation}
    \label{eq:PhaseVelocity}
    \cEM^2 = \frac{1}{\ScalarEpsilon^2}
    = 1 - \frac{\dot{\phi}^2}{M_D^4} .    
\end{equation}
This result is accurate to leading order in $1/M_D$.
The same formula was previously given in Ref.~\cite{Brax:2013nsa}.
In Eq.~\eqref{eq:cp-def} below
we will see this is not precisely the phase velocity associated
with solutions to
the wave equations~\eqref{eq:BWaveEquation}
and~\eqref{eq:BWaveEquation},
although they are related.

Notice that Eq.~\eqref{eq:PhaseVelocity}
is always near unity provided $\dot{\phi}^2 / M_D^4 \ll 1$.
In this calculation we have worked to leading order in $1/M_D$,
so this limitation is effectively a consistency condition.
When $\cEM \ll 1$ higher terms in $1/M_D$ must become important
and we cannot use our calculation to make a clear statement
about the phenomenology.
We describe this as the ``nonlinear region''.
Whether a given model falls in this region depends
explicitly on $M_D$ and the scalar field profile,
and as we will see below this
depends in turn on $\Lambda$ and the expansion
history $H(t)$ of the model.
Unfortunately this nonlinear region is very difficult
to study, because loop corrections will almost
certainly renormalize the coefficients of the
higher-order terms in $1/M_D$.
The functional form of the interaction
is therefore not predictable.
At lowest order this is not a significant concern because
such loop corrections leave the form of the interaction
invariant: their effect can be absorbed into a redefinition
of the scale $M_D$, which is anyway supposed to be unknown.

\para{WKB solution}
Eqs.~\eqref{eq:BWaveEquation} and~\eqref{eq:EWaveEquation}
can be reduced to a common form by making
a field redefinition to remove the friction term
(that is, the term linear in $\dot{\vect{E}}$ or $\dot{\vect{B}}$).
Making the transformations
\begin{equation}
    \vect{E}(t,\vect{x}) = \frac{\vect{G}(t,\vect{x})}{a^{5/2} \ScalarEpsilon^{3/2}} ,
    \qquad \text{and} \qquad
    \vect{B}(t,\vect{x}) = \frac{\vect{G}(t,\vect{x})}{a^{5/2} \ScalarEpsilon^{1/2}}
\end{equation}
it can be checked that both fields can be built from solutions 
to the equation
\begin{subequations}
\begin{equation}
	\label{eq:GWaveEquation}
    \ddot{\vect{G}}
    +
    \meff^2(t) \vect{G}
    -
    \frac{1}{a^2 \ScalarEpsilon^2} \grad^2 \vect{G} = 0 ,
\end{equation}
where the time-dependent mass $\meff^2(t)$ is defined by
\begin{equation}
    \meff^2(t) \equiv
    - \frac{H^2}{4}
    - \frac{\dot{H}}{2}
    - \frac{H}{2} \frac{\ScalarEpsilonDot}{\ScalarEpsilon}
    + \frac{1}{4} \frac{\ScalarEpsilonDot^2}{\ScalarEpsilon^2}
    - \frac{1}{2} \frac{\ScalarEpsilonDDot}{\ScalarEpsilon} .    
\end{equation}
\end{subequations}
Clearly, this merely reflects the fact that both $\vect{E}$
and $\vect{B}$ are derived from the same underlying gauge field.

The terms involving $H^2$ and $\dot{H}$ are generated by mixing
with the metric~\cite{Breitenlohner:1982jf}.
They are both roughly of order $H^2$.
The terms involving derivatives of $\ScalarEpsilon$
will depend on the detailed profile of the scalar field $\phi(t)$.
In Ref.~\cite{Brax:2013nsa} it was suggested that these terms
would also typically be of order $H^2$
unless the field is undergoing a sudden transition.
In~\S\ref{sec:scalar_dynamics} below we will see that this
expectation is borne out for the scalar field profile
generated when the nonlinear kinetic term
$\GalileonOperator{3}$ is relevant.

In this situation we can expect $|\meffDot/\meff^2|$ to be of order
$|\dot{H}/H^2| \sim 1$,
which implies that the mass changes significantly
on timescales of order the Hubble time.
For electromagnetic waves whose wavelength
is much smaller than the horizon we can regard
$\meff^2$
as roughly fixed over many cycles of the wavetrain.
It follows that we can obtain an approximate
description of its evolution using the WKB
procedure.
We expand $\vect{G}$ in a suitable
basis of polarization matrices
with Fourier
mode functions $\psi(t, \vect{k})$.
Then the WKB approximation consists in writing
\begin{equation}
	\label{eq:WKBAnstatz}
	\psi(t, \vect{x}) = \alpha(t)
	\exp
	\Big(
		\im \vect{k}\cdot\vect{x}
		-
		\im \theta(t)
	\Big) ,
\end{equation}
where the phase $\theta(t)$ varies rapidly on the timescale
of $\alpha(t)$.

Substitution of~\eqref{eq:WKBAnstatz}
in~\eqref{eq:GWaveEquation} yields
\begin{equation}
	\frac{\ddot{\alpha}}{\alpha} - \dot{\theta}^2 + \meff^2 + \frac{k^2}{a^2 \ScalarEpsilon^2}
	+
	\im \bigg(
		2 \frac{\dot{\alpha}}{\alpha} \dot{\theta} + \ddot{\theta}
	\bigg)
	= 0 .
\end{equation}
Demanding that the real and imaginary parts cancel separately
shows that the amplitude $\alpha(t)$ varies like
$\alpha(t) = \alpha_0 / \dot{\theta}(t)^{1/2}$.
Meanwhile, the phase function $\theta(t)$ satisfies
\begin{equation}
	\label{eq:WKBPhaseEquation}
	- \frac{1}{2} \frac{\dddot{\theta}}{\dot{\theta}}
	+ \frac{3}{4} \frac{\ddot{\theta}}{\dot{\theta}} \frac{\ddot{\theta}}{\dot{\theta}}
	- \dot{\theta}^2 + \meff^2 + \frac{k^2}{a^2 \ScalarEpsilon^2} = 0 .
\end{equation}
We write
\begin{equation}
	\theta = \thetafast + \thetaslow	 ,
\end{equation}
where $\thetafast$ is designed to absorb the `fast' variation due to integration of the source term,
\begin{equation}
	\label{eq:OmegaEff}
	\frac{\d \thetafast}{\d t} = \bigg( \meff^2 + \frac{k^2}{a^2 \ScalarEpsilon^2} \bigg)^{1/2}
	\equiv \omegaeff(t) ,
\end{equation}
and the last equality defines the effective frequency $\omegaeff$.
The solution is
$\thetafast \approx \int^t \omegaeff(t') \, \d t'$.
In particular,
$\thetafast(t_2) - \thetafast(t_1) \approx \omegaeff(\bar{t}) \Delta t$
if $t_1$ and $t_2$ are not too widely separated,
where $\bar{t} = (t_1 + t_2)/2$ and $\Delta t = t_2 - t_1$.
It follows that the derivative
$\thetafastDot \sim \omegaeff(\bar{t})$ varies much more slowly than $\thetafast$,
making $\thetafastDDot$ a slowly-varying function
that determines $\thetaslow$.
If desired we could solve~\eqref{eq:WKBPhaseEquation}
perturbatively for $\thetaslow$,
although for the purposes of this paper we do not need such precision.
It follows that a typical mode function of the field $\vect{G}$
approximately satisfies
\begin{equation}
	\label{eq:WKBSolution}
	\psi(t, \vect{k})
	=
	\frac{\alpha_0}{\sqrt{\omegaeff(t)}}
	\exp
	\Big(
		\im \vect{k}\cdot\vect{x}
		-
		\im \int^t \omegaeff(t') \, \d t'
	\Big) .
\end{equation}
This solution was previously given in Ref.~\cite{Brax:2013nsa}, neglecting the mass
term $\meff^2$. It was described there as the ``eikonal approximation''.
In this case the WKB approximation reduces to the same procedure.
See also Ref.~\cite{Adshead:2020jqk}, although in this reference the scalar field
is not coupled disformally.

\para{Time-of-flight formula}
The phase velocity of the wavetrain
at time $t$
is roughly
$\omegaeff(t)/k$.
If we
regard an electromagnetic wave as a coherent
superposition of very many collimated photons,
it is clear that the phase velocity of the
wave must equal the propagation velocity of the photons
because of the peculiar properties
of massless particles in special relativity.
For \emph{massive} particles travelling at
less than the speed of light
this relationship is no longer so clear.
In certain circumstances (such as water waves)
it can happen that the phase velocity is
unrelated to the velocity of individual particles
that participate in the wavetrain.

It follows that the phase velocity for~\eqref{eq:WKBSolution}
can be written
\begin{equation}
	c_p
	=
	\frac{\omegaeff}{\kphys}	
	=
	\frac{\cEM}{\kphys}
	\bigg(
		\kphys^2 + \frac{\meff^2}{\cEM^2}
	\bigg)^{1/2}
	=
	\cEM \bigg( 1 + \frac{\meff^2}{\cEM^2 \kphys^2} \bigg)^{1/2}
    \label{eq:cp-def}
\end{equation}
where $\kphys = k/a$ is the physical wavenumber corresponding
to the comoving wavenumber $k$.
Note that the `sound speed' $\cEM$
defined in Eq.~\eqref{eq:PhaseVelocity}
is only equal to the phase velocity if $\meff^2 = 0$.
However,
it is clearly unsatisfactory to regard
the phase velocity $c_p$
as an estimate for the characteristic particle velocity in the beam.
First, $c_p$ can easily become superluminal
even if $\meff^2$ is positive.
Second, there is an unwanted divergence at small $\kphys$,
and at large $\kphys$ (where energies are ultrarelativistic)
$c_p$ approaches $\cEM$ rather than unity.
These properties are entirely characteristic of phase velocities
and have nothing to do with the disformal coupling
or the fact that~\eqref{eq:WKBSolution} is propagating on a curved
background.

Instead, we proceed as follows.
We are still considering~\eqref{eq:WKBSolution}
to describe a coherent superposition of very many collimated
particles
that share an approximate common momentum 4-vector.
Therefore,
consider a small box of spacetime
that lies along the particles' trajectory.
According to the equivalence principle we can
regard~\eqref{eq:WKBSolution} as the wavefunction for an
on-shell particle with 4-momentum $k_\mu$
in the interior of the patch.
For small displacements
$\delta x^\mu = ( \delta t, \delta \vect{x} )$
within the patch this yields
\begin{equation}
	\psi \approx
	\exp
	\Big(
		\im \vect{k} \cdot \delta \vect{x}
		-
		\im \omegaeff \delta t
	\Big)
	\approx
	\e{\im k_\mu \delta x^\mu}
\end{equation}
which implies that we should regard $\omegaeff(t)$
as the characteristic energy for particles in the
beam at time $t$.
This is related to their propagation velocity via
the usual special relativistic formula
$E = \gamma m$, and hence
\begin{equation}
	\label{eq:vDef}
	v = \sqrt{1 - \frac{\meff^2}{\omegaeff^2}} 
	= \cEM
	\bigg(
		\frac{1}{\cEM^2 + \meff^2/\kphys^2}
	\bigg)^{1/2} .
\end{equation}
Clearly $v$ has more satisfactory properties than $c_p$.
As $\kphys \rightarrow 0$ the propagation velocity approaches zero.
In the ultrarelativistic limit $\kphys \rightarrow \infty$
we have $v \uparrow 1$.
Further, $v$ is always subluminal if $\meff^2$ is positive.

Using $v$ to estimate particle velocities in the beam, the
time of flight between two locations $A$ and $B$
is
\begin{equation}
	\label{eq:TimeOfFlight}
	T(A \rightarrow B) = \int_A^B \frac{\d r}{v} ,	
\end{equation}
where $\d r$ is an element of length along the
trajectory
and
we have assumed that $A$ and $B$ are sufficiently local
that the effects of curvature can be ignored.
This is typically the case for LIGO sources.
For local sources it will also be a reasonable
approximation to take $v$ as time independent,
in which case the travel time of electromagnetic
radiation $\TEM$
compared to the travel time of gravitational radiation
$\Tgrav$ will be
\begin{equation}
    \label{eq:TimeOfFlight-approx}
	\TEM \approx \frac{1}{v} \Tgrav	.
\end{equation}
To obtain a quantitative estimate requires
information about the scalar field profile.
We discuss this in~\S\ref{sec:scalar_dynamics}
before applying~\eqref{eq:TimeOfFlight}--\eqref{eq:TimeOfFlight-approx}
to GW170817 in~\S\ref{sec:conclusion}.

\section{The dynamics of the scalar field}\label{sec:scalar_dynamics}

Our task is now to solve for the evolution $\phi(t)$ of the Galileon scalar.
To leading order in $1/M_D$ the action can be obtained by
linearizing Eq.~\eqref{equation1}
in the disformal coupling.
Explicitly, this is
\begin{equation}
	\label{eq:ExplicitScalarAction}
	\int \d^4x \; \sqrt{-g} \,
	\bigg[
		\frac{\Mp^2}{2} R-\LambdaCC
	    + \frac{c_{2}}{2} \dot{\phi}^{2} 
	    +\frac{1}{\Lambda^3} \ddot{\phi} \dot{\phi}^{2} 
	    + \bigg( 1-\frac{\dot{\phi}^2}{2M_D^4} \bigg)\rho
	\bigg] ,
\end{equation}
where $\rho$ is the density of baryonic and cold dark matter.
Eq.~\eqref{eq:ExplicitScalarAction}
applies at any epoch, but we mostly use it during matter
domination where $\rho = \rho_m$.
As explained in~\S\ref{sec:intro},
we fix $c_2$ in order to break a scaling symmetry in
the Galileon sector;
its role is to make the quadratic kinetic term
\emph{individually}
stable if $c_2 = +1$
and ghostlike if $c_2 = -1$.
We will allow the cubic self-interaction scale
$\Lambda$ to vary over a suitable parameter range.
It can be positive or negative.

Remarkably, Eq.~\eqref{eq:ExplicitScalarAction}
admits an exact solution in which the
linear and nonlinear $\phi$ kinetic
terms combine to support a nontrivial field profile.
Applying the Euler--Lagrange equation to~\eqref{eq:ExplicitScalarAction}
in terms of $\dot{\phi}$
requires
$\partial{\mathcal L}/\partial\dot{\phi}= \d/ \d t \big(\partial{\mathcal L}/\partial\ddot{\phi}\big)$.
This leads immediately to the \emph{algebraic} solution
\begin{align}
	\label{eq:ScalarProfile}
    \dot{\phi}(t) = -\bigg(c_2+\frac{\rho_m(t)}{M_D^4}\bigg)\frac{\Lambda^3}{3H(t)} .
\end{align}
In the absence of the nonlinear term the only solution
available is $\dot{\phi} = c$ for constant $c$.
This is substantially less interesting and does not lead
to time-dependent effects from the disformal
coupling in Eqs.~\eqref{eq:BWaveEquation}--\eqref{eq:EWaveEquation}
or~\eqref{eq:GWaveEquation}.
In the late universe $\rho_m$ is negligible in comparison with
$M_D$ once we impose the ATLAS constraint
$M_D \gtrsim 1.2 \, \TeV$~\cite{Aaboud:2019yqu}.
Therefore we see that the disformal coupling
cannot play an important role in the evolution of $\phi$
except during the very early universe, before the time of the
electroweak phase transition.

Eq.~\eqref{eq:ScalarProfile} is valid for all cosmological backgrounds
and parameter values, provided that $\dot{\phi}^2 \ll M_D^4$.
In the late universe its time variation is set by $H(t)$,
as indicated in~\S\ref{sec:EMWaves}.
Notice that this solution bears a strong resemblance
to the ``tracker''
solutions described by Barreira et al.~\cite{Barreira13,Renk:2017rzu},
which are defined so that
\begin{equation} \label{tracker_solution_eq}
    \frac{\dot{\phi} H}{\Mp H_{0}^{2}} \equiv \xi= \text{constant} .
\end{equation}
These solutions yield $\dot{\phi} \sim 1/H$,
as does~\eqref{eq:ScalarProfile}
when the disformal coupling can be neglected.

\para{Parameter constraints}
Recall that to evade stringent constraints from the ISW effect
we do not allow the Galileon to contribute significantly to the energy budget
of the Universe.
The energy density contributed by the Galileon sector is
\begin{equation} \label{gal_density}
    \rhoGal
    =
    \rhoTwo
    +
    \rhoThree
    =
    \frac{1}{2} c_2 \dot{\phi}^{2}
    + \frac{1}{\Lambda^3} \ddot{\phi} \dot{\phi}^{2} ,
\end{equation}
where $\rhoTwo$
and $\rhoThree$ measure the energy density contributed
by the quadratic and cubic Galileon operators
$\GalileonOperator{2}$ and $\GalileonOperator{3}$, respectively.
We constrain the model parameters
so that $\rhoGal$
evaluated at the present day
is smaller than the current background energy
density $\sim \meV^4$.
In practice
this roughly requires $|\Lambda| \lesssim 10^{-12} \, \eV$
independent of $M_D$.
In terms of the solution~\eqref{eq:ScalarProfile}
we can evaluate
$\rhoTwo$ and $\rhoThree$ individually,
\begin{align}
    \rhoTwo
    & \equiv
    \frac{1}{2} c_2\dot{\phi}^2
    \approx
    \frac{1}{2} c_2\frac{\Lambda^6}{9H^2} ,
    \\
    \rhoThree
    & \equiv
    \frac{1}{\Lambda^3} \ddot{\phi}\dot{\phi}^2
    =
    \frac{2}{3} \frac{\dot{H}}{H^2}
    \bigg(
    	c_2
    	+ \frac{\rho_m}{M_D^4}
    	- 6\frac{H^2}{M_D^4}
    \bigg)c_2 \rhoTwo
    \approx
    \frac{c_2}{3}
    \frac{\dot{H}}{H^2}\frac{\Lambda^6}{9H^2} .
\end{align}
In the final expressions we have assumed $\rho_m \ll M_D^4$,
which will generally be the case except at very early times.

This is not the only constraint.
We also require the Galileon sector to be ghost-free
and to be free of Laplacian instabilities.
Explicit formulae for these conditions were given
in the Jordan frame
by Appleby \& Linder~\cite{Appleby:2011aa}.
In principle these
conditions should be corrected due to
the Einstein-frame disformal coupling in~\eqref{equation1},
but in practice these corrections are not numerically important
because of the condition $\dot{\phi}^2 / M_D^4 \ll 1$.
When evaluated on the background~\eqref{eq:ScalarProfile},
the no-ghost condition requires
\begin{equation}
    \label{eq:no-ghost}
    \frac{3 c_2}{2} - \frac{c_2^2 \Lambda^6}{27 \Mp^2 H^4} < 0 .
\end{equation}
The second term is roughly proportional to
$\rhoGal / \rho_m$ is therefore small
whenever the Galileon energy density is subdominant.
It follows that the stable sector of the theory
has $c_2 = -1$.%
    \footnote{This differs from the condition $c_2=+1$
    that would be required for stability of the quadratic
    term by itself.
    The reason is that on the background solution~\eqref{eq:ScalarProfile},
    both the quadratic and cubic operators are relevant.}
This is consistent with Appleby \& Linder's observation
that absence of ghosts typically requires
the Galileon energy density to be negative.

\begin{figure}[h!] \label{avoid_catast}
\centering
\includegraphics[scale=0.9]{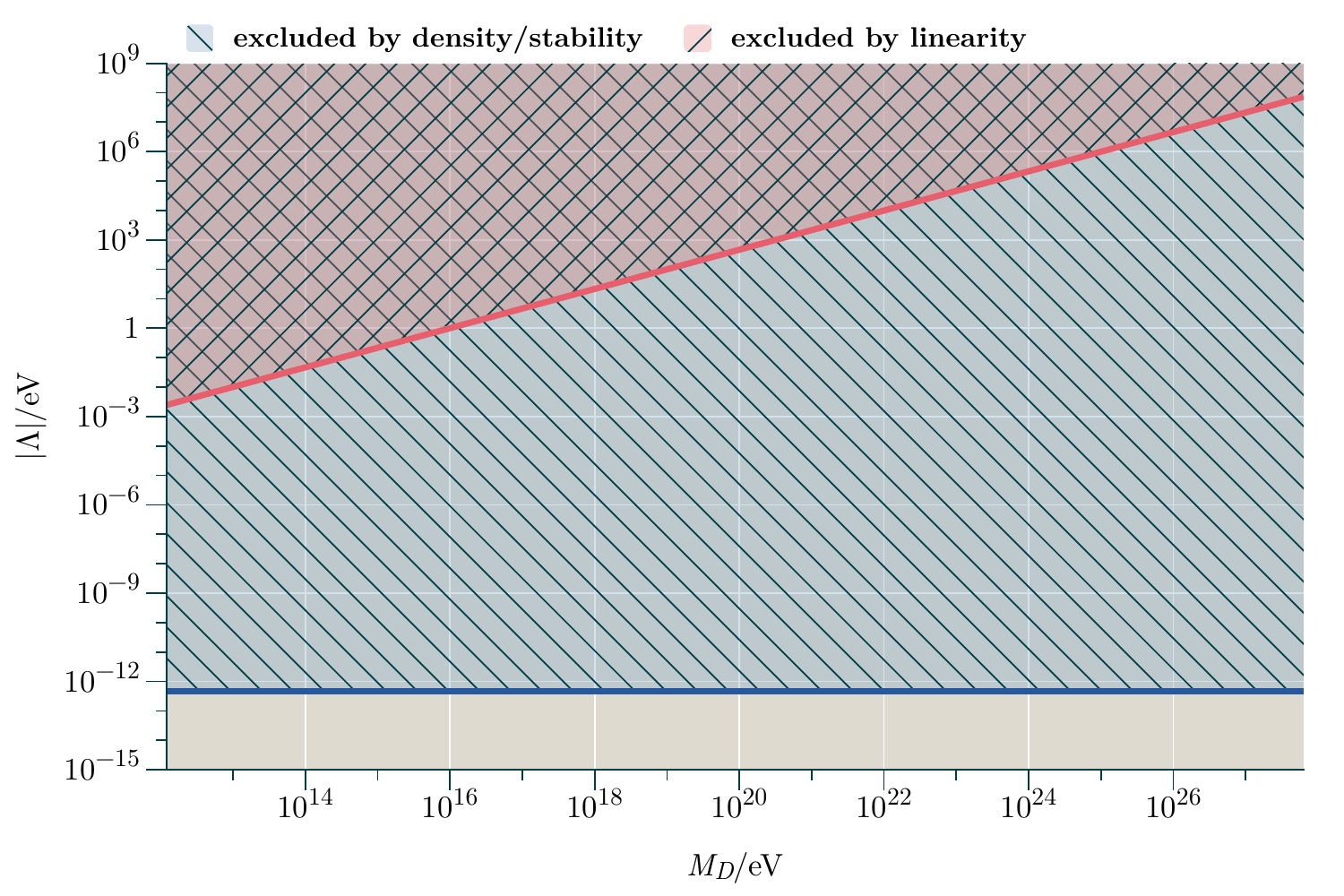}
\caption{Excluded parameter space for the cubic Galileon model
with disformal
coupling to matter.
The density/stability constraint
(blue) should be considered uncertain by a factor of
order unity
that determines the maximum
amplitude of $\rhoGal$ relative to $\rho_m$.
Eqs.~\eqref{eq:no-ghost}
shows that a similar order unity factor
determines how closely we approach the
boundary of the ghost-like region.
The red region is excluded
because the solution approaches the
`nonlinear' region where $\dot{\phi}^2/M_D^4$ is
no longer negligible.}
\label{figure}
\end{figure}

Under the same approximations,
assuming $c_2 = -1$
and matter domination,
the Laplacian constraint
requires
\begin{equation}
    \label{eq:laplacian-stability}
    \left( 1 + \frac{8 H_0}{H} \right)
    -
    \frac{8 (3H + 2 H_0) \Lambda^6}{729 \Mp^2 H^5}
    > 0 .
\end{equation}
This is also satisfied automatically
provided the Galileon energy density is
subdominant.
Therefore the no-ghost and Laplacian stability constraints
do not generate independent limits on $|\Lambda|$
beyond those already imposed by $\rhoGal$.
We plot these in Fig.~\ref{figure}.

\section{Conclusion}\label{sec:conclusion}

We are now in a position to apply the time-of-flight
formula~\eqref{eq:TimeOfFlight}
to GW170817.
Assuming~\eqref{eq:ScalarProfile}
for the scalar field profile,
we find that the difference in travel time
is negligible,
despite the tightness of the constraint.
We
assume the lower limit for $M_D$
allowed by collider measurements~\cite{Aaboud:2019yqu},
and take
$\Lambda = 2.4 \times 10^{-13} \, \eV$,
which is the largest value permitted
by the constraints on $\rhoGal$
for this value of $M_D$.
These choices maximize the time-of-flight difference.
Unfortunately,
for any physically reasonable $\kphys$,
it can be checked that $\TEM$ and $\Tgrav$
are indistinguishable to more than 15 significant
figures
for any low-redshift event such as GW170817.
Therefore we conclude that the disformal coupling
\emph{alone}
is so weak
it cannot be constrained even by
precise time-of-flight observations
on the background~\eqref{eq:ScalarProfile}.
A similar conclusion applies to
the dispersion of light
implied by the $k$-dependence
of~\eqref{eq:WKBSolution}
and~\eqref{eq:vDef}.

Notice that there is a curious competition between
$\rhoTwo$ and $\rhoThree$
during matter domination.
Because the linear kinetic energy density is proportional
to $\dot{\phi}^2$, the constraint is independent of the sign of $\Lambda$.
It is also almost independent of the sign of
$c_2$.%
	\footnote{A discussion of the frame dependence is
	considered in Ref.~\cite{Faraoni:1998qx, Flanagan:2004bz}.}
Prior to dark energy domination the linear and non-linear terms
$\rhoTwo$ and $\rhoThree$
in Eq.~\eqref{gal_density} have almost exactly the same amplitude;
their ratio is $c_2 2\dot{H}/(3H^2)\simeq c_2$ during matter domination,
up to corrections of order $\rho_m/M_D^4$. Hence,
on the stable branch where $c_2=-1$, the
total scalar kinetic energy is always suppressed during matter domination
until dark energy dominates at $z\lesssim 0.5$.
This `screening-like' effect
is curious and is worthy of further attention.
Today, for a flat $\Lambda$CDM
cosmology with $\Omega_m\simeq0.29$ the linear energy dominates by a
factor of 3.5 over the non-linear term.

Our analysis supports earlier impressions that a disformal coupling
is difficult to constrain using cosmology alone.
If so, then collider physics will remain the best prospect for
determining constraints on the disformal coupling scale $M_D$,
but conversely we cannot expect the current ATLAS
bound $M_D \gtrsim 1.2 \, \TeV$ to be dramatically superseded
in the near- or medium-term future.
While previous studies including a Galileon sector and
a disformal coupling have reported that best-fit models
typically yield a time-lag $\Delta t$ too large to be compatible
with LIGO/Virgo constraints, the effect in these analyses
is driven by the internal structure of the Galileon
sector and not by the disformal coupling.

\section*{Acknowledgements}
MGL acknowledges support from the UK Science and Technology Facilities Council
via Research Training Grant ST/M503836/1.
CB and DS acknowledge support from the Science and Technology Facilities Council [grant number ST/T000473/1].

\appendix

%\begin{thebibliography}{99}
\bibliographystyle{apsrev}%{aipauth4-1}
\bibliography{bib}

% Please avoid comments such as "For a review'', "For some examples",
% "and references therein" or move them in the text. In general,
% please leave only references in the bibliography and move all
% accessory text in footnotes.

% Also, please have only one work for each \bibitem.

%\end{thebibliography}
\end{document}